\documentclass[10pt,preprint]{aastex}
\usepackage{amsmath}
\usepackage{graphicx}
\usepackage{graphics}
\usepackage[dvips]{color}
\newcommand{\kms}{{km~s$^{-1}$}}

\begin{document}

\title{Reddening and \ion{He}{1}* $\lambda10830$ Absorption Lines in Three Narrow-line Seyfert 1 Galaxies}
\author{Shaohua Zhang\altaffilmark{1}, Hongyan Zhou\altaffilmark{1,2}, Xiheng Shi\altaffilmark{1}, Wenjuan Liu\altaffilmark{3},
Xiang Pan\altaffilmark{1,2}, Ning Jiang\altaffilmark{2}, Tuo Ji\altaffilmark{1}, Peng Jiang\altaffilmark{1}, Shufen Wang\altaffilmark{1,2}}
\altaffiltext{1}{Polar Research Institute of China, 451 Jinqiao Road, Shanghai, 200136, China; zhangshaohua@pric.org.cn; zhouhongyan@pric.org.cn}
\altaffiltext{2}{Key Laboratory for Researches in Galaxies and Cosmology, Department of Astronomy, University of Sciences and Technology of China, Chinese Academy of Sciences, Hefei, Anhui, 230026, China}
\altaffiltext{3}{Key Laboratory for the Structure and Evolution of Celestial Objects, Yunnan Observatories, Chinese Academy of Sciences, Kunming, Yunnan 650011, China}

\begin{abstract}
We report the detection of heavy reddening and the \ion{He}{1}* $\lambda10830$ absorption lines at the AGNs' redshift in three Narrow-line Seyfert 1 galaxies: SDSS J091848.61+211717.0, SDSS J111354.66+124439.0, and SDSS J122749.13+321458.9.
They exhibit very red optical to near-infrared colors,  narrow Balmer/Paschen broad emission lines and \ion{He}{1}* $\lambda10830$ absorption lines.
The ultraviolet-optical-infrared nucleus continua are reddened by the SMC extinction law of $E(B-V)\sim 0.74$, 1.17, and 1.24 mag for three objects, which are highly consistent with the values obtained from the broad-line Balmer decrements, but larger than those of narrow emission lines. The reddening analysis suggests the extinction dust simultaneously obscure the accretion disk, the broad emission line region, and the hot dust from the inner edge of the torus. It is possible that the dust obscuring the AGN structures are the dusty torus itself. Furthermore, the Cloudy analysis of the \ion{He}{1}* $\lambda10830$ absorption lines propose the distance of the absorption materials to be the extend scale of the torus. That greatly increases probabilities of the obscure and absorption materials being the dusty torus.
\end{abstract}

% (SDSS J091848.61+211717.0, SDSS J111354.66+124439.0, SDSS J122749.13+321458.9)}

\section{Introduction}
Active galactic nuclei (AGNs) emit over the entire electromagnetic spectrum, and are widely believed to be produced by the accretion of matter onto a supermassive black hole (SMBH) in the center of galaxy. Narrow-line Seyfert 1 galaxies (NLS1s) are the special subclass of AGNs in the evolution history. NLS1s typically show the narrower Balmer lines ($FWHM < 2000 ~\rm km~s^{-1}$), the stronger \ion{Fe}{2} emission ($\rm Fe~II/H\beta >0.5$), the weaker [\ion{O}{3}] lines ($\rm [O~III]/H\beta < 3$), the stronger variability and steeper X-ray spectra (e.g., Goodrich 1989; Boroson \& Green 1992; Wang, Brinkmann, \& Bergeron 1996;  Leighly 1999; Marziani et al. 2001; Grupe et al. 2004) than other AGNs. NLS1s  have lower black hole masses and higher $L/L_{\rm Edd}$, thus they are considered to be in the quick growth stage of the central black hole (e.g., Mathur 2000; Grupe \& Mathur 2004; Zhou et al. 2006; Komossa 2008; Xu et al. 2012).

Because of these properties, NLS1s are generally thought to be in the early evolution stage (Grupe et al. 1999), and perhaps link between AGNs and normal galaxies (Zhang et al. 2009; Rovilos et al. 2009). According to the current AGN unified model, the differences between NLS1s and other Seyfert galaxies, i.e., broad-line Seyfert 1s and Seyfert 2s, are ascribed solely to orientation effects and anisotropic obscuration (Antonucci 1993). Although many studies suggest that NLS1s are observed from a nearly pole-on view (e.g., Osterbrock \& Pogge 1985; Puchnarewicz et al. 1992; Taniguchi et al. 1999), it becomes apparent that they cannot be easily brought into the frame of the unified model.
Fortunately, partially obscured objects provide a new way to see the inside of AGNs (e.g., Dong et al. 2005). Through the studies of the SED extinction and the associated absorption lines, they are expected to provide some of the constraints on the physical and geometrical conditions in the centers of AGNs.

In this work, we report the extremely reddening and the detection of \ion{He}{1}* $\lambda10830$ absorption lines at the AGNs' redshift in three NLS1s: SDSS J091848.61+211717.0, SDSS J111354.66+124439.0, and SDSS J122749.13+321458.9 (hereafter SDSS J0918+2117, SDSS J1113+1244, and SDSS J1227+3214), which were initially reported by the red Two Micron All Sky Survey (2MASS; Skrutskie et al. 2006) AGNs (Kuraszkiewicz et al. 2009) and the FIRST-2MASS red quasars (F2M; Glikman et al. 2007; 2012). We picked them up again during the series study of the reddening AGNs (e.g., Li et al. 2015; Liu et al. 2016; Pan et al. 2017; Zhang et al. 2017). We study the \ion{He}{1}* $\lambda10830$ absorption lines (the first detection reported by Leighly et al. 2011) as well as the continuum, and hydrogen Balmer/Paschen broad emission lines of these objects. Different from the dust located on the inside of the torus in another NLS1 SDSS J2339-0919 (Yang et al. in prepare), the heavy reddening and the \ion{He}{1}* $\lambda10830$ absorption lines in these three objects are due to the dust from the torus itself.

\section{Multiwavelength Observations}
SDSS J0918+2117, SDSS J1113+1244, and SDSS J1227+3214 were best known for infrared-luminous dust-reddened quasars, and have been comprehensively observed. For these three objects, the broad-band spectral energy distributions (SEDs) from the far-ultraviolet (FUV) out to the middle-infrared (MIR) are presented by the photometric data taken with the Galaxy Evolution Explorer (GALEX; Morrissey et al. 2007), the Sloan Digital Sky Survey (SDSS; York et al. 2000), the 2MASS survey, and the Wide-field Infrared Survey Explorer (WISE; Wright et al. 2010). Three objects in the SDSS images are unresolved and are classified to `STAR', `psfMag' is theoretically the optimal measure of their brightness. However, since the redshifts of SDSS J0918+2117 and SDSS J1227+3214 are 0.1493 and 0.1368 (Hewett \& Wild 2010), we choose Petrosian magnitudes for two objects.\footnote{Details can be seen in  http://classic.sdss.org/dr6/algorithms/photometry.html\#which.} Meanwhile, they are included in the 2MASS All-Sky Point Source Catalog (PSC), rather than the Extended Source Catalog (XSC) for their unresolved images.
Their optical to near-infrared colors suggest that they are  much more heavily reddened than the normal red quasars and less than the true type-2 quasars.The photometric magnitudes are summarized in Table \ref{tab1}.

Their optical spectra of were taken with the SDSS 2.5 m telescope on 2004 March to 2006 March.
The SDSS spectrographs, which provide high signal-to-noise ratio (SNR) spectra at the resolution
$R \sim 1800$ and the wavelength coverage from 3800 to 9200 \AA\ (Stoughton et al. 2002).
We access the 1-D spectra of them from the SDSS spectroscopic catalog Data Release Seven (DR7; Abazajian et al. 2009).
Meanwhile, the Hubble Space Telescope (HST) Space Telescope Imaging Spectrograph (STIS) archive
provides a low-resolution UV spectrum for SDSS J0918+2117 (Kuraszkiewicz et al. 2009).
The HST/STIS archival data can make the spectral blueward of SDSS J0918+2117 effectively reach to $\sim 1700$ \AA\ in the rest-frame.

Through the Chinese Telescope Access Program (TAP), their NIR spectra were performed with the TripleSpec spectrograph
on the Hale 200-inch telescope (P200) at Palomar Observatory on 2014 January and 2015 March.
Four exposures of 300, 90 and 240 seconds each (for SDSS J0918+2117, SDSS J1113+1244, and SDSS J1227+3214, respectively)
are taken in an A-B-B-A dithering model. TripleSpec (Wilson et al. 2004) provides simultaneous wavelength coverage from 0.9 to 2.46 $\mu$m at a resolution of 1.4 - 2.9 \AA, with two gaps at approximately 1.35 and 1.85 $\mu$m owing to the telluric absorption bands.
The raw data were processed using Interactive Data Language (IDL) based Spextool software (Vacca et al. 2003; Cushing et al. 2004).
The TripleSpec spectra detect   \ion{He}{1}$*~\lambda10830$ absorption lines and P$\alpha$  emission lines of SDSS J0918+2117 and SDSS J1227+3214, and  H$\alpha$ emission line of SDSS J1113+1244.

Furthermore, SDSS 1113+1244 was followed up by Urrutia et al. (2008), as one of F2M red quasars with the the Advanced Camera for Surveys (ACS) Wide Field Camera on HST (GO-10412). It was imaged with the $F475W$ and the $F814W$ filters at 2005 May 26. The point-source subtraction and host galaxy fitting of the HST/ACS images will help to confirm the SED analysis.

\section{Broadband Spectral Energy Distribution}

\begin{figure*}[tbp]
\epsscale{1.0} \plotone{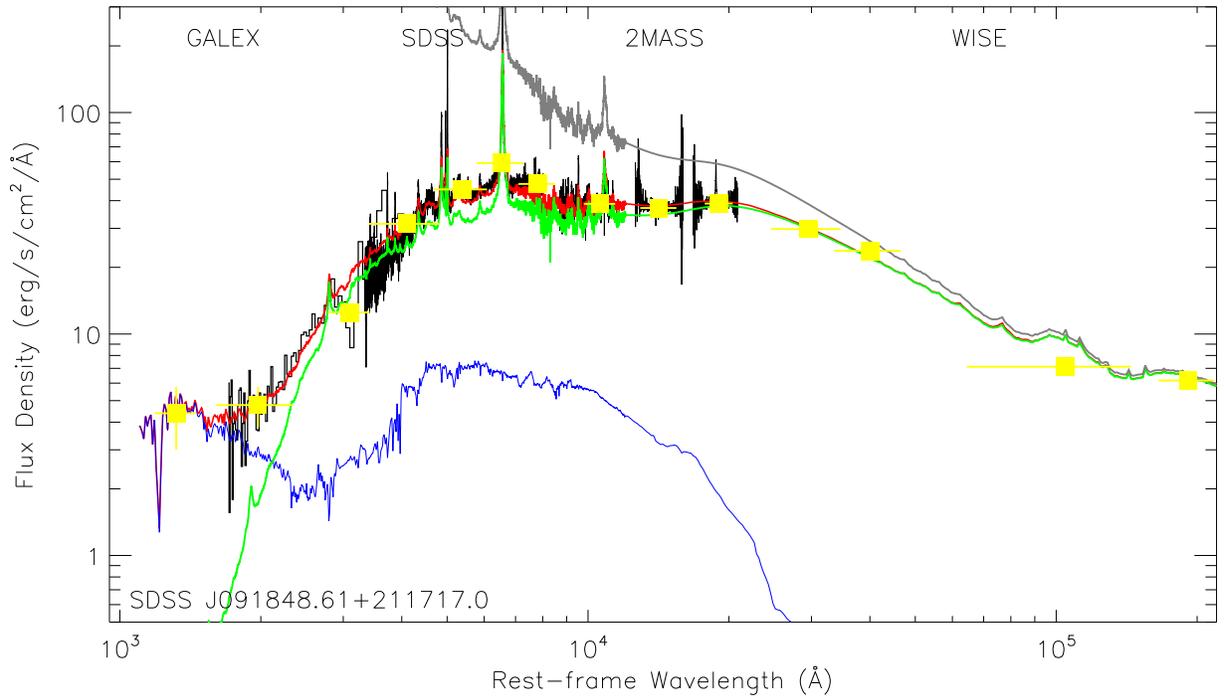}
\caption{Broadband SED of SDSS~J0918+2117 from FUV to MIR by yellow squares, the spectra of HST/STIS, SDSS and TripleSpec by black curves. The reddened quasar composite with $E(B-V)=0.74\pm0.01$, the Sc galaxy template and their sum are shown by green, blue and red curves. Overplotted for comparison is the quasar composite (gray curve).
}\label{f1}
\end{figure*}

\begin{figure*}[tbp]
\epsscale{1.0} \plotone{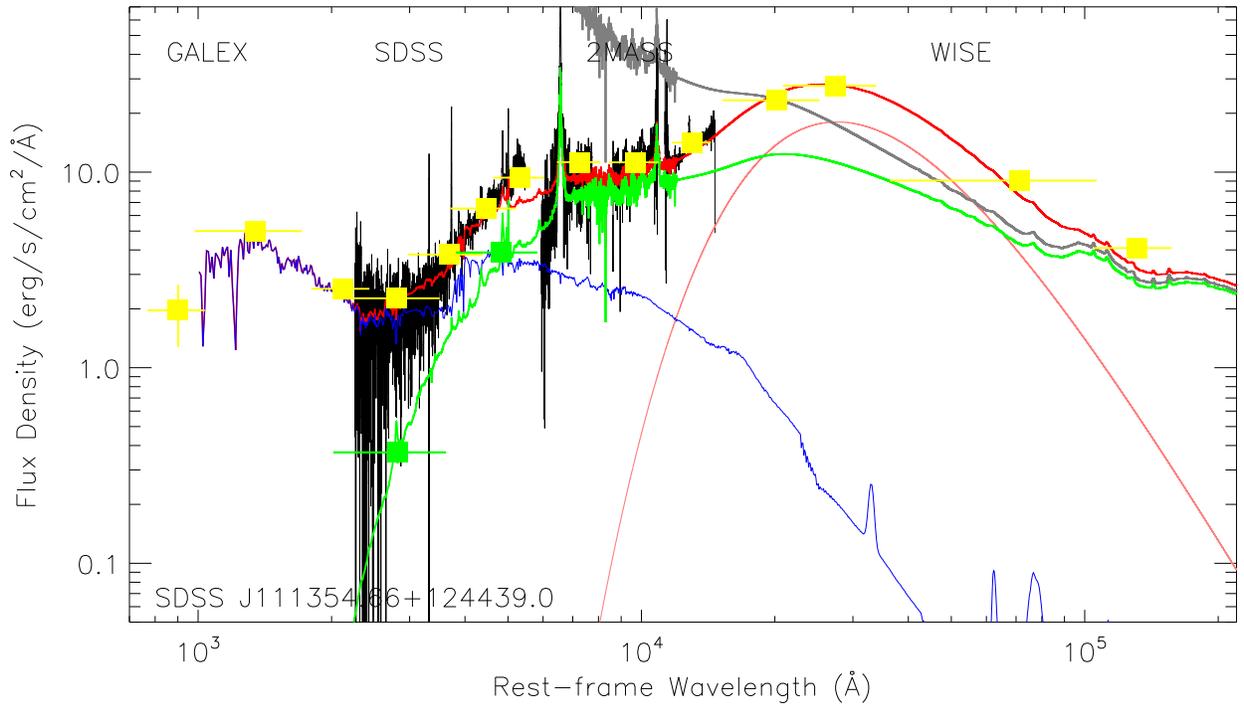}
\caption{Broadband SED of SDSS~J1113+1244 from FUV to MIR by yellow squares, the spectra of SDSS and TripleSpec by black curves.
The  reddened quasar composite with $E(B-V)=1.17\pm0.02$, the Sd galaxy template, the warm dust emission with $T \sim 1050$ K,
and their sum are shown by green, blue, pink and red curves. Overplotted for comparison is the quasar composite (gray curve).
}\label{f2}
\end{figure*}

\begin{figure*}[tbp]
\epsscale{1.0} \plotone{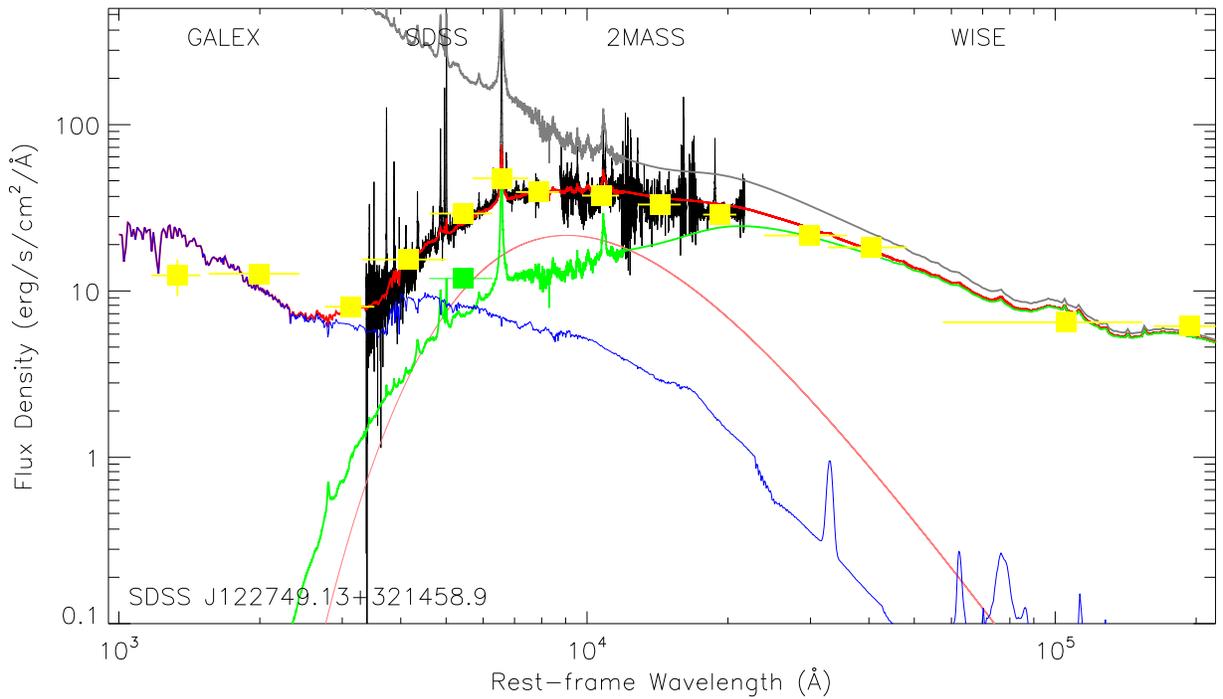}
\caption{Broadband SED of SDSS~J1227+3214 from FUV to MIR by yellow squares, the spectra of SDSS and TripleSpec by black curves.
The reddened quasar composite with $E(B-V)=1.24\pm0.02$, the Sdm galaxy template,
the high-temperature ($T\sim  3200$ K) black-body emission, %the M1 star spectrum,
and their sum are shown by green, blue, pink and red curves.
Overplotted for comparison are the quasar composite (gray curve) and the GALFIT PSF Magnitude from the SDSS $r-$band image (green square).
 }\label{f3}
\end{figure*}

After correcting for the Galactic reddening using the dust map of Schlegel et al. (1998) and the Fitzpatrick (1999) reddening curve, we transformed the multiwavelength photometric data and the spectroscopic fluxes into the rest-frame with the redshifts ($z \rm =0.1493,~ 0.6805,~and~ 0.1368$; Hewett \& Wild 2010). Meanwhile, the  UV, optical and NIR spectra are scaled to match the GALEX, SDSS and 2MASS photometry. % for the UV, optical and NIR spectra.
The photometric and spectroscopic data are shown by yellow squares and black curves in Figures \ref{f1}, \ref{f2}, and \ref{f3}.
Generally, NLS1s exhibit significant variability on timescales from about 10 days to a few years with smaller amplitudes on average compared to Broad-line Seyfert 1 galaxies (BLS1s), and the optical/UV variability is wavelength dependent¡ªthe shorter the wavelength and the larger the variation s(e.g., Ai et al. 2013).
However, in this work, the time ranges used to compile the multi-wavelength SEDs covering a wavelength range from the FUV to the mid-IR are spanning 10 to 15 years in the rest frame, variability of the continuum strength and shape could be an issue. Fortunately, the two facts, i.e., the perfect matching of the optical/NIR spectra simply multiplied by a factor and the broad-band SEDs, and the high consistency of extinctions from the SED fitting and the Balmer decrements (see next section),  rule out the possibility of variability affecting on the SED in these three NLS1s.
When we take a overview of the features of the SEDs, it is obvious that the observed SEDs of all the three objects show very different shapes from that of the average quasar spectrum, but in contrast to be similar with the heavily obscured quasar SDSS J000610.67+121501.2 reported in Zhang et al. (2017). In the optical, and NIR bands, they is much lower than the average quasar spectrum; from the $u$- or $NUV$-band up to lower wavelengths, the SEDs slightly turn up (extremely weak in SDSS J0918+2117). The unique shape is considered to be caused by the spectral combination of two different continuum slopes (Zhang et al. 2017). When looking at the emission/absorption lines, the optical and NIR spectra show the significant broad/narrow emission lines, \ion{He}{1}$^*~\lambda 10830$ and weak \ion{Ca}{2} H and K (3969 and 3934 \AA) absorption lines, which represent the contributions of the AGN and its host galaxy.

To model the SEDs, we used a phenomenological model due to the lack of a good physical understanding of the nature and geometry of the AGN and its host galaxy. Two principal components are superposed into the  model: (1) the quasar composite and (2) the host galaxy starlight.
The quasar composite (also used in Liu et al. 2016) is created by splicing the UV to optical quasar composite of Vanden Berk et al. (2001), the NIR quasar composite of Glikman et al. (2006), and the FIR quasar composite of Netzer et al. (2007). It contains the nucleus power-law continuum, the broad/narrow emission lines from the broad/narrow line regions, the NIR and MIR emission from the torus, and the more or less starlight contamination.
We used the galaxy templates in the SWIRE library (Polletta et al. 2007) to approximately match the broad-band SED shape of the host galaxy contribution. The SWIRE library provides us the galaxy templates of 6 starbursts, 3 ellipticals, and 7 spirals (ranging from early to late types, S0-Sdm) to represent the host galaxy starlight, which have a enough wavelength coverage of \mbox{0.1 - 1000 $\mu$m}. In addition, a warm dust component is added for SDSS J1113+1244 to make up the NIR peak located at the WISE $W2$-band.
For SDSS J1227+3214, we add an extra high-temperature black-body emission to repair the mismatch between the  model and observed spectrum. Details will be given in the discussion of the individual objects below.
Finally, we can decompose the broad-band SED in rest-frame wavelength with the following model:
\begin{align}
F_{\lambda}=C_1 A_{E(B-V),\lambda} F_{\lambda,\rm QSO} + C_2 F_{\lambda,\rm HOST} +C_3 B_{\lambda}(\rm T_{dust}),
\label{eq1}
\end{align}
where $F_{\lambda}$ is the observed spectrum, $F_{\lambda,\rm QSO}$, $F_{\lambda,\rm HOST}$,
and $B_{\lambda}(\rm T_{dust})$ are the quasar composite, the galaxy template,
and the Planck function, respectively;
$C_x~(x=1, 2,\rm and~3)$ are the factors for the the respective components; $A_{E(B-V),\lambda}$ is the dust extinction to the  quasar composite by the SMC extinction law, and no reddening were assumed for other spectral components. The last item is alternative as needed,
and the F-test analysis will decide whether need to add the Planck function for the potential warm dust or other unknown component.
We perform least-squares minimization using the IDL procedure MPFIT developed by Markwardt (2009).

SDSS J0918+2117 --  The combination of a scaled and reddened quasar composite with $E(B-V)=0.74\pm0.01$ and a scaled Sc galaxy template can closely follow the observed magnitudes and spectra from the FUV-band to the $W4$-band. The sum and components are shown by red, and green and blue curves in Figure \ref{f1}. In Kuraszkiewicz et al. (2009),
the effects of host galaxy on AGN optical/NIR colors suggest the host-galaxy of SDSS J0918+2117 is an Sa galaxy with a 5 Gyr stellar population, and the host galaxy strength is 20 times weaker than the intrinsic/unreddened AGN at $R$-band (Sa05,20). In contrast, our modeled Sc template contributes approximate 14\% of total observed flux at the same band, in general agreement to the estimation of Kuraszkiewicz et al. considering of the extinction correction. But for the constraint of the GALEX magnitudes, we will obtain the same host galaxy contribution (Sa05,20) as Kuraszkiewicz et al. (2009).

SDSS J1113+1244 -- The best-fitting results are shown in Figure \ref{f2}. We recovered the same $E(B -V ) = 1.17$ value as reported in  Urrutia et al. (2008), which appears well-fit by this model. And the starlight template of the host galaxy also is Sd galaxy. However, a  dust emission with $T \sim 1050$ K (pink curve) is added as suggested in Urrutia et al. (2012), which was interpreted as the extra dust close to the sublimation radius (also see Glikman et al. (2006), Netzer et al. (2007) and Mor \& Trakhtenbrot (2011)). However, the typical dust sublimation temperature is $\sim 1500-2000$ K (e.g., Tuthill et al. 2001; Monnier \& Millan-Gabet 2002), the extra `warm' dust in this object is probably not the inner edge of the torus.
Its HST/ACS images show the clear stellar-like feature (the AGN component) with the PSF Magnitudes of $22.87 \pm 0.25$ and $19.11 \pm 0.05$ for the F475W and the F814W filters (Table 5 of Urrutia et al. 2008; green squares), which have a good agreement with our modeled quasar component (green curve).

SDSS J1227+3214 -- Its phenomenological model is similar with that of SDSS J1113+1244, the only difference is the mismatch of the two-component model and the observed data at the optical band. A reddened quasar composite and an Sdm galaxy template can be almostly determined by the GALEX, SDSS-$u$ and WISE photometric (green and blue curves in Figure \ref{f3}). The mismatch (with the center wavelength of $\sim 8000$ \AA) looks like a black-body emission with a high temperature of $\sim  3200$ K (pink curve in Figure \ref{f3}). That suggests that there are more low-mass (e.g., M1-type) stars in the host galaxy. The final three-component model is consistent with the UV-to-IR observed magnitudes. Addition of the high temperature black-body indeed makes the phenomenological model look weird, thus, we tried to enhance the host dust emission to make up the mismatch. However, the hot ($\sim 1500-2000$ K) dust at the sublimation radius can not shift the peak of dust emission to the mismatch wavelength range.
We also tried to perform a two-dimensional decomposition of the AGN and host galaxy for the $r-$band image using GALFIT (Peng et al. 2010),
in which the image is adopted as the PSF+S\'{e}rsic model. An effective decomposition is expected to put an independent constraint on the SED of the AGN. While the GALFIT shows that the $r-$band image contains the PSF Magnitude of 18.35 (green squares for comparison), the S\'{e}rsic component of 18.22 mag, $n=2.81$, and $r_e=0.62''$, under great uncertainty, since the SDSS image has a relatively short exposure time and low spatial resolution due to the seeing limit. Like SDSS J1113+1244, the HST/ACS imaging in the future can seek final clarification. The host galaxy starlight (Sdm template and perhaps  low-mass stars) is the dominant component of the UV and optical wavelength band. The modeled  starlight of SDSS J1227+3214 lowers the the contribution of the AGN, thus raises our measured $E(B-V)$ from 0.71 mag and 0.94 mag (LaMassa et al. 2016; Glikman et al. 2012) to $1.24\pm0.02$ mag. On the other hand, the Balmer decrement analysis in the next section shows that the extinction to the Balmer lines is highly consistent with that to the broad-band SED, the consistence strengthens our confidence about the phenomenological model of SDSS J1227+3214.

To summarize these results, unitary extinction can resolve the AGN's reddening from the UV out to the MIR by the SMC extinction law, that suggests that the dust simultaneously affect the UV-optical continuum and the IR emission of the torus, in other words, the dust  obscure the AGN structures inside of the torus. It is possible that they are the torus itself or the dust exterior to the torus. Meanwhile, studies of the host galaxies of NLS1s suggest that NLS1s tend to have smaller host galaxies (e.g., Krongold et al. 2001), and the host galaxies of NLS1s show a higher fraction of bars (e.g., Crenshaw et al. 2003; Ohta et al. 2007; Bian \& Huang 2010; Olgu{\'{\i}}n-Iglesias et al. 2017), grand-design nuclear dust spirals and stellar nuclear rings (e.g., Deo et al. 2006; Le{\'o}n Tavares et al. 2014). Similarly, the SED fitting shows the host galaxies of these three NLS1s are all the late type spirals, which might indicate more efficient fueling of their black holes.  %( Sc, Sd, and Sdm galaxies, respectively).}

\section{Balmer and Paschen Emission Lines}

As shown in Figures \ref{f1}, \ref{f2}, and \ref{f3}, broad emission lines of hydrogen Balmer and Paschen series were detected by the SDSS optical spectrographs and the P200 TripleSpec spectrograph, with the only exception that in SDSS J1113+1244, P$\alpha$ line lies beyond the wavelength coverage of the TripleSpec, due to the higher systemic redshift.
Through the measurement of Balmer and Paschen broad lines and the comparison with the intrinsic flux ratios of AGNs, the dust extinction can be derived from the observed Balmer decrements at a different angle.

\begin{figure*}[tbp]
\epsscale{1.0} \plotone{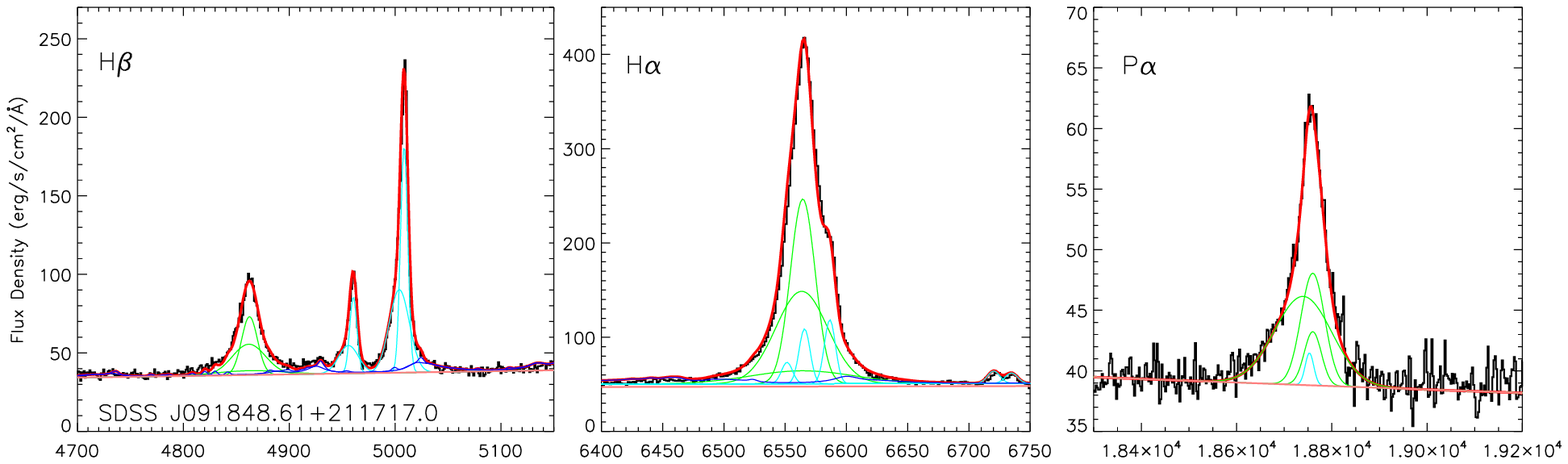}\plotone{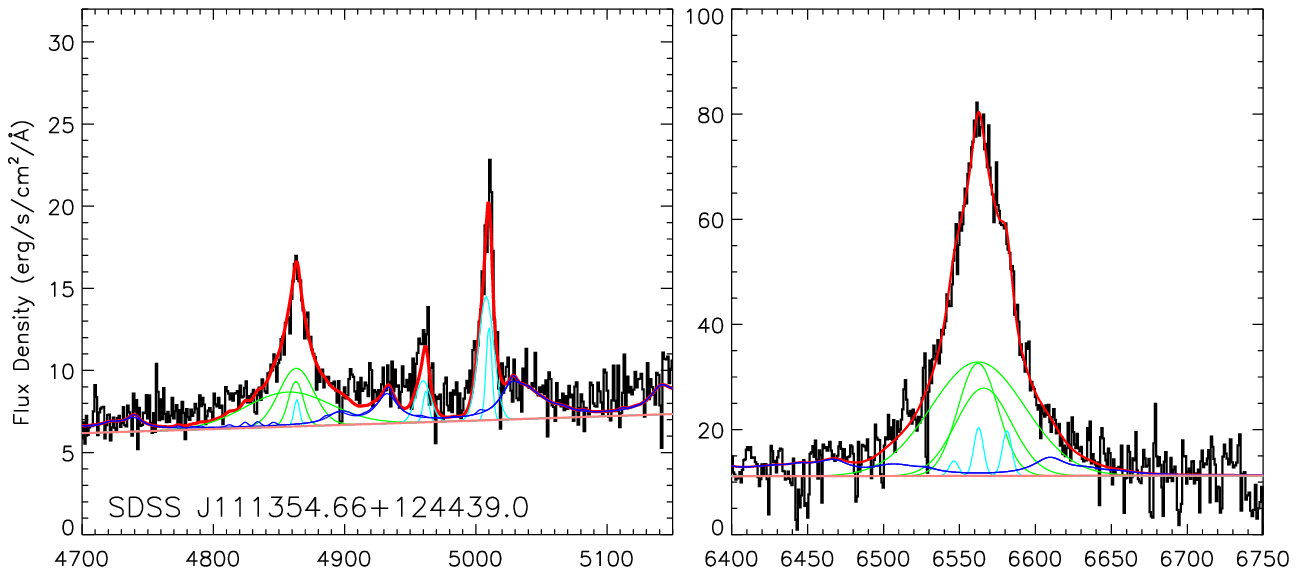}\plotone{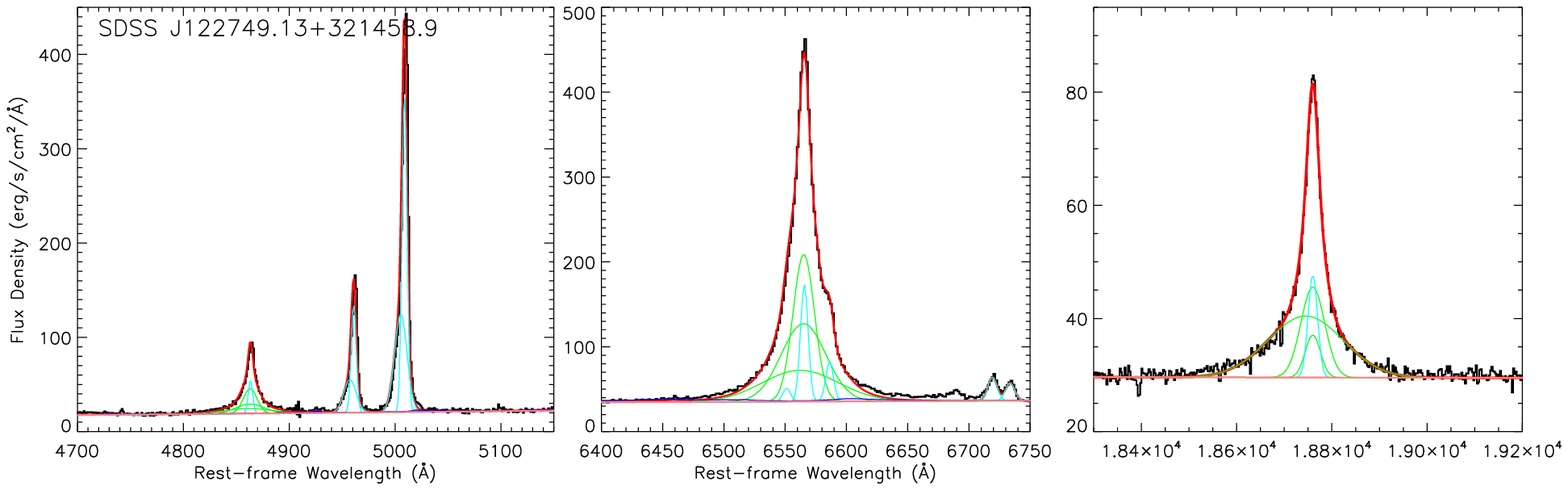}
\caption{Observed spectra of H$\beta$, H$\alpha$ and P$\alpha$ regimes with the best-fit models. The models are described in detail in Section 4. In each panel, we plot the observed spectrum in black curves, first-order polynomial continuum in pink, broadened broad/narrow \ion{Fe}{2} multiplets in blue, Gaussian broad components in green, Gaussian narrow components in cyan and the model sum in red.
}\label{f4}
\end{figure*}

In this work, we are only concerned about strong and easily decomposable  lines, i.e., H$\alpha$, H$\beta$, and P$\alpha$,
and ignored other complicated and/or low-SNR lines.
The optical and NIR spectra are heavily reddened and contaminated by the host galaxy, as introduced above,
thus the power-law usually used is hard to describe the continuum in a large wavelength range.
For the three spectral regimes: H$\beta$ + [\ion{O}{3}] $\lambda\lambda$4959,5007,
H$\alpha$ + [\ion{N}{2}] $\lambda\lambda$6548,6583, and P$\alpha$,
a first-order polynomial is adopted to finetune the local continuum (independently for each regime) within a limited wavelength range.
Because the Balmer lines are heavily blended with strong \ion{Fe}{2} multiplets,
in the H$\beta$ + [\ion{O}{3}] $\lambda\lambda$4959,5007 and H$\alpha$ + [\ion{N}{2}]  $\lambda\lambda$6548,6583 regimes, we also adopted the \mbox{I Zw 1} \ion{Fe}{2} template provided by V{\'e}ron-Cetty et al. (2004) and convolved it with a Gaussian kernel in velocity space to match the widths of broad/narrow \ion{Fe}{2} multiplets in the observed spectra. The polynomial and \ion{Fe}{2} multiplets were collectively referred to as the underlying pseudocontinuum, and were firstly subtracted from the observed spectra to further study the emission lines.
The continuum windows are [4200, 4720] \AA\ and [5080, 5300] \AA\ for H$\beta$ + [\ion{O}{3}] $\lambda\lambda$4959,5007, [6150, 6250] \AA\ and
[6850, 6950] \AA\ for H$\alpha$ + [\ion{N}{2}]  $\lambda\lambda$6548,6583, and [1.82, 1.84] $\mu$m and [1.92, 1.94] $\mu$m for P$\alpha$. It is worthy noting that the broad emission lines (listed in Table 2 of Vanden Berk et al. 2001) in the continuum windows are masked in the pseudocontinuum fitting. In the panels of Figure \ref{f4}, the local continua and broad/narrow \ion{Fe}{2} multiplets are marked by pink and blue lines.

After subtracting the pseudocontinuum models, we modeled the broad and narrow mission lines as multiple Gaussians: three Gaussians for broad Balmer and Paschen components, and one Guassian for narrow Balmer and Paschen components and other narrow lines.
Additionally, one extra Gaussian was adopted to model the blue wing of [\ion{O}{3}] line (e.g., Komossa et al. 2008; Zhang et al. 2011).
The [\ion{O}{3}] $\lambda\lambda$4959,5007 doublets were assumed to have the same shift  and profile; the same was applied to [\ion{N}{2}] $\lambda\lambda$6548,6583 doublets and [\ion{S}{2}] $\lambda\lambda$6716,6731 doublets (e.g., Osterbrock \& Pogge 1987). The flux ratios of the  [\ion{O}{3}] doublets and [\ion{N}{2}] doublets were fixed to the theoretical values of 2.98 (e.g., Dimitrijevic et al. 2007) and 2.96 (Storey \& Zeippen 2000), respectively.
The measured [\ion{S}{2}] $\lambda6716/\lambda 6731$ ratios have a large dispersion, from 0.81 to 1.44 (Table 5 in Osterbrock \& Pogge 1987), then the flux ratio of the  [\ion{S}{2}] doublet is free in the fitting.
The details of these components: broad components in green and narrow components in cyan, are shown in the panels of Figure \ref{f4}. The measured broad line parameters are listed in Table \ref{tab2}.

In Veilleux and Osterbrock (1987), they adopted $\rm H\alpha/H\beta=3.1$ for active galaxies and 2.85 for \ion{H}{2} region galaxies (e.g., Ferland \& Netzer 1983; Gaskell 1984; Gaskell \& Ferland 1984). Recently, the large sample statistics suggests that the intrinsic value of broad-line $\rm H\alpha/H\beta$ is 3.06 with a standard deviation of 0.03 dex (Dong et al. 2008). Gaskell \& Ferland (1984) presented the relative strengths of P$\alpha$ and H$\beta$ for low-density gas was  very close to the Case-B value, 0.34, here it was assumed to be applicable for the broad emission lines.
Based on the observed Balmer decrements of $\rm H\alpha/H\beta$ and $\rm P\alpha/H\beta$, we calculated the values of $E(B-V)$ assuming the extinction curve of the SMC (Table \ref{tab2}). The deviation of the intrinsic ratios did not enter into our calculation, otherwise, larger errors will be applied to the estimated $E(B-V)$. The extinctions to the Balmer lines are highly consistent with those to the broad-band SEDs,
the broad emission line regions (BLRs) of these three objects have the same intense dust-enshrouded as the continuum.
While the measurements from the Balmer decrement $\rm P\alpha/H\beta$ is currently smaller, it is possible that the problem lies in the intrinsic  relative ratio of P$\alpha$ and H$\beta$ broad lines.

\begin{figure*}[tbp]
\epsscale{0.6} \plotone{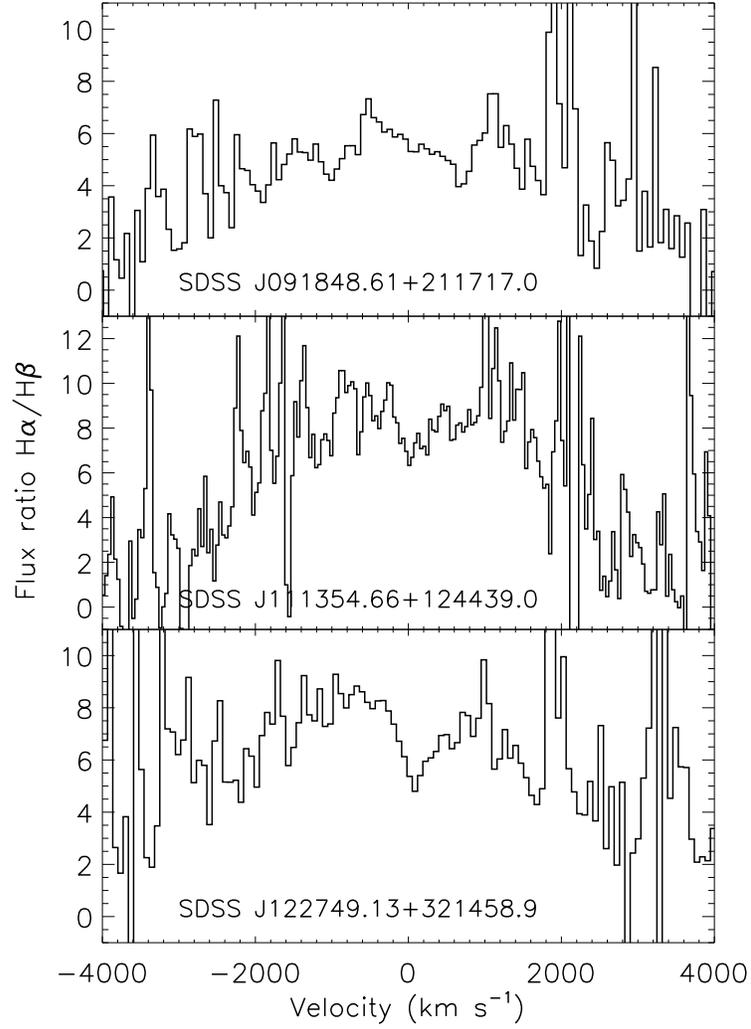}
\caption{Observed profile ratios of H$\alpha$ and H$\beta$ emission lines in velocity space. The troughs at 0 \kms\ in the profile ratio curves imply that the Balmer narrow lines have smaller extinction than the Balmer broad lines.}\label{f5}
\end{figure*}

For NLS1s, the strength and decrement measurements of the narrow emission lines are heavily dependent on the emission-line profile decomposition and have large uncertainty, thus we do not employ the results of narrow lines and list them in Table \ref{tab2}. To study the extinction difference of broad and narrow emission lines, we try to use the profile ratio of Balmer lines to explore the difference. In Figure \ref{f5}, we show the flux ratios of H$\alpha$ and H$\beta$ emission lines in the velocity space, where the emission profiles are calculated from the observed spectra subtracting the underlying pseudocontinuum and  [\ion{N}{2}]  $\lambda\lambda$6548,6583 doublet. It is clear that there are the  obvious  troughs at 0 \kms\ in the profile ratio curves, which implies the narrow emission line region (NLR) has smaller extinction than the BLR. Generally, the dust in the AGNs exist in the  large-scale galactic environment or the dusty torus around the AGN. The extinction difference  of broad and narrow emission
lines suggests that the dust affecting on the SED and the BLR are from the dusty torus.

\section{\ion{He}{1}*$~\lambda10830$ Absorption Lines}
\begin{figure*}[tbp]
\epsscale{1.0} \plotone{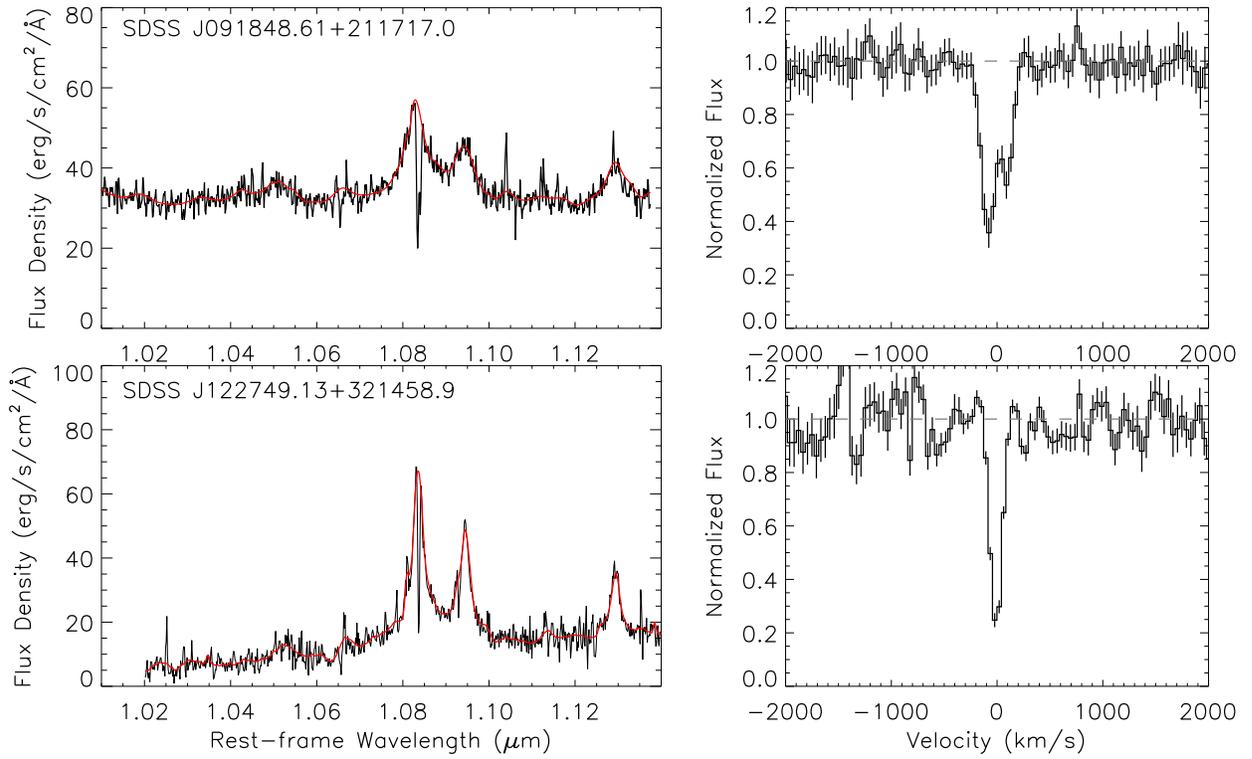}
\caption{Left: Observed spectra subtracting the host galaxy contribution (black) and unabsorbed fluxes from the spline interpolation (red) of \ion{He}{1} $\lambda10830$ regimes. Right: Normalized spectra of \ion{He}{1}* $\lambda 10830$ absorption lines in the velocity space.
}\label{f6}
\end{figure*}

The redshifted \ion{He}{1}$~\lambda10830$ emission lines of two objects (SDSS J0918+2117 and SDSS J1227+3214) are detected by the $J$-band TripleSpec spectra Figures \ref{f1} and \ref{f3}), and the narrow ($\sim$ 300 - 400 \kms) and deep ($\sim$ 60 \%) \ion{He}{1}*$~\lambda10830$ absorption troughs are identified in the peaks of the emission lines. Unfortunately, the $H$-band TripleSpec spectrum only covers part of the blue wing of \ion{He}{1}$~\lambda10830$ emission line of SDSS J1113+1244 with low-SNR, and the absorption region of \ion{He}{1}*$~\lambda10830$ falls into the gap at 1.85 $\mu$m. The same extinction of the reddening quasar SEDs and Balmer decrements suggests the absorption materials can obscure the power-law continuum, broad emission lines, and the hot dust emission, rather than the host galaxy. Thus, when we calculated the normalized spectra, we firstly removed  the host galaxy contribution (blue (and pink) curves in Figures \ref{f1} and \ref{f3}) from the observed spectra. Since the \ion{He}{1}*$~\lambda10830$ absorption troughs are quite narrow, the unabsorbed fluxes are estimated using spline interpolation by masking the absorption troughs (Left panels of Figure \ref{f6}). In right panels, we also plot the normalized spectra of \ion{He}{1}*$~\lambda10830$ absorption troughs in the velocity space obtained by dividing the observed spectra subtracting the host galaxy starlight by the unabsorbed fluxes.
Note that, the typical $1\,\sigma$ error of the starlight is $\sim 10\%$, the effect of the starlight uncertainties to the normalized spectra is negligible.

Theoretically, for a partially obscured absorber, the normalized intensity in the troughs is
\begin{align}
I(v)=1-Cf(v)+Cf(v)e^{-\tau(v)},
\label{eq3}
\end{align}
where $Cf(v)$ is the percentage covering factor of the absorber, and $\tau(v)$ is the optical depth.
% for the relevant ion. Using the optical depths derived above,
Then, the column densities of  \ion{He}{1}$^*$ as a function of velocity are calculated using the
general expression (e.g., Arav et al. 2001)
\begin{align}
N(\Delta v)&=\frac{m_e c}{\pi e^2}\frac{1}{\lambda_0 f_0}\tau(\Delta v) \nonumber\\
            &= \frac{3.7679\times10^{14}}{\lambda_0 f_0} \tau(\Delta v)~\rm [cm^{-2}~(km~s^{-1})^{-1}],
\label{eq4}
\end{align}
where $\lambda_0=10830\rm ~\AA$ and $f_0=0.5392$ are the wavelength and the oscillator strength\footnote{Oscillator strengths are from NIST Atomic Spectra Database (http://physics.nist.gov/PhysRefData/ASD/)} of \ion{He}{1}*$~\lambda10830$, respectively. The total column density  is obtained  by integrating  Equation \ref{eq4}.

For SDSS J0918+2117 and SDSS J1227+3214, the minimum residual fluxes in the absorption troughs are 0.36 and 0.25, respectively. There are two extreme possibilities: (1) The \ion{He}{1}*$~\lambda10830$ absorption line is completely obscured but unsaturated, the column density of \ion{He}{1}* can be directly calculated to be $N_{\rm HeI*} = (1.44\pm0.27)\times10^{13}~ \rm cm^{-2}$ for SDSS J0918+2117, and $N_{\rm HeI*} = (1.08\pm0.09)\times10^{13}~ \rm cm^{-2}$ for SDSS J1227+3214. Since the optical depth ratio of \ion{He}{1}*$~\lambda10830/\lambda3889 = 23.1$, the \ion{He}{1}*$~\lambda 3889$ absorption trough is so weak that the trough will be drowned in the spectral noise. %This is in agreement with the observation.
(2) The \ion{He}{1}*$~\lambda10830$ absorption line is partly  obscured but saturated. In this situation, we used  the maximum absorption depths of the \ion{He}{1}*$~\lambda10830$ troughs as the covering factors for two cases, which are $Cf = 0.64$ and 0.75, respectively. Due to the spectral noises within the potential \ion{He}{1}*$~\lambda 3889$ absorption troughs, the upper limits of \ion{He}{1}*$~\lambda 3889$ absorption depths are 0.13 and 0.20 at $2\,\sigma$ level. Then the upper limits of the \ion{He}{1}* column densities in SDSS J0918+2117 and SDSS J1227+3214 are approximated to be $N_{\rm HeI*} = 5.48\times10^{13}~ \rm cm^{-2}$  and $4.40 \times 10^{13}~ \rm cm^{-2}$, respectively.
Both above situations are in agreement with the observation.
The true $N_{\rm HeI*}$ of the \ion{He}{1}*$~\lambda10830$ absorption lines in SDSS J0918+2117 and SDSS J1227+3214 should be somewhere in-between them.

As we known, the column density of \ion{He}{1}* is insensitive to hydrogen density $n_{\rm H}$ in a wide range,
while H(n=2) is sensitive to the density of the gas; meanwhile, \ion{He}{1}* grows in front of the ionization front of
hydrogen and stops growing behind it, i.e., $N_{\rm HeI*}$ depends on the ionization parameter $U$
(e.g., Arav et al. 2001; Ji et al. 2015; Liu et al. 2015; Sun et al. 2017).
Thus, the absorptions of metastable helium and hydrogen Balmer are the good indicators of $U$ and $n_{\rm H}$, respectively.
Furthermore, the ionization parameter depends on the distance $R$ away from the central source and the ionization photonemission rate $Q$, as follows,
\begin{eqnarray}
U =\int^{inf}_{\nu_0} \dfrac{L_{\nu}}{4\pi R^2 h\nu n_{\rm H}c}~d\nu=\dfrac{Q}{4\pi R^2 n_{\rm H}c},
\end{eqnarray}
where $\nu_0$ is the frequency corresponding the hydrogen edge and $c$ is the speed of light.
We scaled the ionizing continuum to the extinction corrected continuum flux at 5100 \AA\ and then integrate over the hydrogen edge of this scaled SED.
This yields $Q = 1.62 \times 10^{55}~\rm and ~1.14 \times 10^{55}~ \rm photons~s^{-1}$ for SDSS J0918+2117 and SDSS J1227+3214, respectively.
Using the derived $Q$, $U$, and $n_{\rm H}$, the value of $R$ will be inferred.

\begin{figure*}[tbp]
\epsscale{1.} \plotone{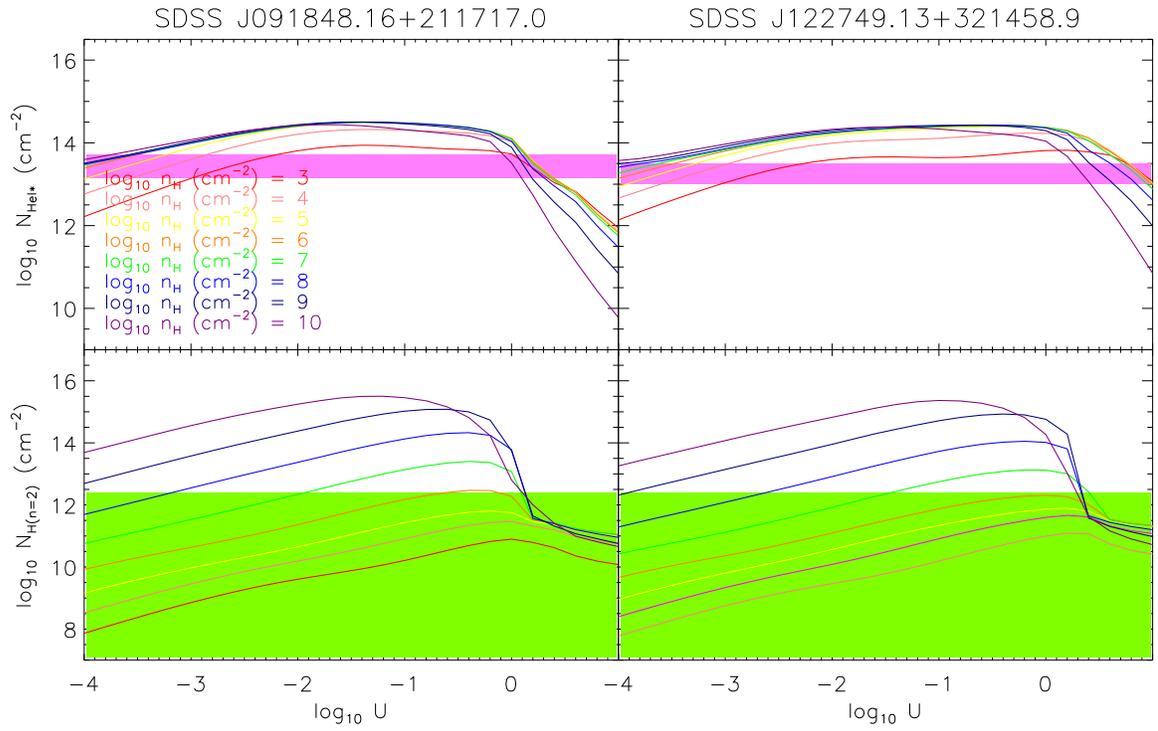}
\caption{The column densities of the \ion{He}{1}*  and H(n=2) ions as functions of ionization parameter $U$ for electron densities from $10^3$ to $10^{10}~\rm cm^{-3}$. The light magenta and light green areas are the estimated $N_{\rm HeI*}$ and $N_{\rm H(n=2)}$ ranges, respectively. }\label{f7}
\end{figure*}

The photoionization synthesis code CLOUDY (the latest version, last described by Ferland et al. 1998) is applied to simulate the ionization process. We considered a slab-shaped geometry, unique density, homogeneous chemical composition of solar values, and a SED of ionizing continuum of commonly used the typical AGN multicomponent continuum.\footnote{See details in Hazy, a brief introduction to Cloudy; http://www.nublado.org.} This UV-soft SED is regarded as more realistic for radio-quiet quasars than the other available SEDs provided by Cloudy (see the detailed discussion in Section 4.2 of Dunn et al. 2010).
The above analysis shows the absorption material is a mixture of dust grains and gas, and thus the dust-to-gas ratio ($E{\rm ( B-V)}/N_{\rm H}$) and the depletion of various elements from the gas phase into dust should be taken into account in the models.
The values of $E{\rm ( B-V)}$ are given from the extinction, and on account of the results of the X-ray spectral analysis, the hydrogen column densities of two absorbers are fixed to $\rm log_{10} ~N_{\rm H}~({\rm cm^{-2}}) = 21.6$ %in the following simulation
(SDSS J0918+2117: $N_{\rm H} \sim 4 \times 10^{21} \rm ~ cm^{-2}$, Pounds \& Wilkes 2007; SDSS J1227+3214: $N_{\rm H} = 3.4^{+0.8}_{-0.7} \times  10^{21}~\rm cm^{-2}$, LaMassa et al. 2016).
We calculated a series of photoionization models with different ionization parameters and gas densities. The ranges of parameters are $-4\leqslant{\rm log_{10}}~U\leqslant 1$ and $3\leqslant{\rm log_{10}}~n_{\rm H}~({\rm cm^{-3}})\leqslant10$ with a step of 0.2 dex. Figure \ref{f7} shows the column densities of the \ion{He}{1}* and H(n=2) ions as functions of ionization parameter $U$. The light magenta areas are the estimated $N_{\rm HeI*}$ ranges of SDSS J0918+2117 and SDSS J1227+3214 (top panels). The overlapping regions are two possible ionization parameter spaces (in front or behind the ionization front) for the \ion{He}{1}*$~\lambda10830$ absorption. We also estimated the upper limits of H$\alpha$ absorption depths of 0.10 and 0.26 at $2\,\sigma$ level from the spectral noises. If H$\alpha$ absorption lines have the similar profiles of \ion{He}{1}*$~\lambda10830$, the corresponding column densities are $N_{\rm H(n=2)}= 2.44 \times 10^{12} \rm~cm^{-2}$ and $3.55 \times 10^{12} \rm~cm^{-2}$, those constrain the gas density to be in the light green areas (bottom panels). For the ionization parameter region of $\rm log_{10} ~U \gtrsim 0$, the extremely high ionization can only exist in the BLR of the AGNs (Figure 5.3 in Peterson 1997),
meanwhile, the dust also can not survive in the such high ionization environment.
For the low-ionization region, we carefully calculated the ranges of $U$ for a given $n_{\rm H}$, and found the lower limit of $R$ with the parameter combination of $\rm log_{10} ~n_{\rm H}~({\rm cm^{-3}}) = 8$ and $\rm log_{10} ~U\sim -3.5$.
The lower limits of $R$ are $\sim 9.5$ pc and 8.2 pc for SDSS J0918+2117 and SDSS J1227+3214, respectively.
Similarly, the extend scales of the torus are on the scale of $\sim 10$ pc (Burtscher et al. 2013; Kishimoto et al. 2011).
The analysis of absorption lines also suggest the absorption materials are perhaps the dusty torus itself.

\section{CONCLUSION}
We performed a multiwavelength study of the continuum, Balmer and Paschen broad emission lines and \ion{He}{1}* $\lambda10830$ absorption lines  of three NLS1s. Their optical-infrared colors present the extremely reddening of the continuum, thus they were initially reported as red 2MASS AGNs or F2M quasars. The multiwavelength SED analysis indicates that the nucleus continua of SDSS J0918+2117, SDSS J1113+1244, and SDSS J1227+3214 are heavily suppressed by dust reddening with $E (B-V ) = 0.74\pm 0.01$, $1.17\pm0.02$, and $1.24\pm0.02$, respectively. When we use the galaxy templates in the SWIRE library to represent the host galaxy contribution, the host galaxy types are Sc, Sd, and Sdm galaxies, and the galaxy starlights dominant the emission of the UV-band.
The follow-up TripleSpec observations on the P200 telescope provide H$\alpha$/P$\alpha$ emission lines and \ion{He}{1}* $\lambda10830$ absorption lines. The emission line fitting suggests that the extinctions to the Balmer broad lines are $E(B-V)=0.83\pm0.04$, $1.15\pm0.07$, and $1.30\pm0.04$, highly consistent with those to the broad-band SEDs.
Meanwhile, the profile ratio of Balmer lines in the velocity space show the NLR has smaller extinction than the BLR.
SDSS J0918+2117 and SDSS J1227+3214 have measured $E{\rm (B-V)}/N_{\rm H}$ values of $1.85\times10^{-22}$ \mbox{mag cm$^2$} and $3.65\times10^{-22}$ \mbox{mag cm$^2$} on account of the hydrogen column densities of the X-ray spectral analysis.
The dust to gas ratios in two cases are 8.5 and 17 times higher than that of the SMC (Weingartner \& Draine 2001).
Based on measurements of \ion{He}{1}* $\lambda10830$ absorption lines, the  \ion{He}{1}* column densities are limited to the small range of several $10^{13}$ cm$^{-2}$. Combined with the hydrogen column densities and the undetected Balmer absorption lines, the photoionization simulation with dust finds the lower limit ($\sim \rm 9.5~pc~ and ~8.2~ pc$) of the absorption materials away from the center source with the parameter combination of $n_{\rm H} = 10^8 \rm~cm^{-3}$ and $U=10^{-3.5}$, which are the extend scale of the torus. The analysis of absorption lines confirm the dust location inferred from the reddening again.
The obscure and absorption materials are perhaps the dusty torus itself.

\acknowledgments
We are thankful for the helpful comments from an anonymous referee on this paper.
This work is supported by the National Natural Science Foundation of China (NSFC-11573024, 11473025, 11421303), and the National Basic Research Program of China (2013CB834905). We acknowledge the use of the Hale 200-inch Telescope at Palomar Observatory through the Telescope Access Program (TAP), as well as the archive data from the GALEX, SDSS, 2MASS, and WISE surveys. TAP is funded by the Strategic Priority Research Program, the Emergence of Cosmological Structures (XDB09000000), National Astronomical Observatories, Chinese Academy of Sciences, and the Special Fund for Astronomy from the Ministry of Finance. Observations obtained with the Hale Telescope at Palomar Observatory were obtained as part of an agreement between the National Astronomical Observatories, Chinese Academy of Sciences, and the California Institute of Technology. Funding for SDSS-III has been provided by the Alfred P. Sloan Foundation, the Participating Institutions, the National Science Foundation, and the U.S. Department of Energy Office of Science. The SDSS-III Web site is http://www.sdss3.org/

\clearpage

\begin{deluxetable}{cc lcc ccc ccc cc}
\tabletypesize{\scriptsize}
\tablewidth{0pt}
\rotate
\tablenum{1}
\tablecaption{Multiwavelength Observations
\label{tab1} }
\tablehead{
\colhead{} & \colhead{} &\multicolumn{2}{c}{SDSS J0918+2117} & \colhead{} & \multicolumn{2}{c}{SDSS J1113+1244} & \colhead{} &\multicolumn{2}{c}{SDSS J1227+3214} & \colhead{} & \colhead{} & \colhead{}\\
\cline{3-4} \cline{6-7} \cline{9-10}\\
\colhead{Band} &\colhead{} & \colhead{Magnitude} & \colhead{Obs. Date }& \colhead{} &\colhead{Magnitude} & \colhead{Obs. Date } &\colhead{} & \colhead{Magnitude} & \colhead{Obs. Date }& \colhead{} & \colhead{Survey/Telescope} & \colhead{References}}
\startdata
$FUV$        &  &$22.88\pm 0.61^a$ &2010 Feb. 04& &$ 23.61\pm 0.47 $ & 2008 Apr. 02 && $ 21.60\pm 0.31 $ & 2007 Apr. 14&  &GALEX &1\\
$NUV$        &  &$ 21.95\pm 0.25 $ &2010 Feb. 04& &$ 21.74\pm 0.12 $ & 2008 Apr. 02 && $ 20.72\pm 0.17 $ & 2007 Apr. 14&  &GALEX &1\\
\textit{u} &  &$ 19.82\pm 0.03 $ &2004 Dec. 13& &$ 21.55\pm 0.15 $ & 2008 Nov. 02 && $ 20.18\pm 0.05 $ & 2008 Nov. 02&  &SDSS&2,3\\
\textit{g} &  &$ 18.17\pm 0.01 $ &2004 Dec. 13& &$ 21.04\pm 0.04 $ & 2003 Mar. 31 && $ 18.79\pm 0.01 $ & 2004 May 12 &  &SDSS&2,3\\
\textit{r} &  &$ 17.13\pm 0.01 $ &2004 Dec. 13& &$ 19.87\pm 0.02 $ & 2003 Mar. 31 && $ 17.53\pm 0.01 $ & 2004 May 12 &  &SDSS&2,3\\
\textit{i} &  &$ 16.42\pm 0.01 $ &2004 Dec. 13& &$ 18.85\pm 0.01 $ & 2003 Mar. 31 && $ 16.60\pm 0.01 $ & 2004 May 12 &  &SDSS&2,3\\
\textit{z} &  &$ 16.25\pm 0.01 $ &2004 Dec. 13& &$ 18.05\pm 0.02 $ & 2003 Mar. 31 && $ 16.41\pm 0.01 $ & 2004 May 12 & &SDSS&2,3\\
$  J  $      &  &$ 14.81\pm 0.04 $ &1998 Jan. 12& &$ 16.14\pm 0.08 $ & 1997 Dec. 18 && $ 14.83\pm 0.04 $ & 2000 Apr. 12& &2MASS&4\\
$  H  $      &  &$ 13.74\pm 0.01 $ &1998 Jan. 12& &$ 15.03\pm 0.07 $ & 1997 Dec. 18 && $ 13.85\pm 0.04 $ & 2000 Apr. 12& &2MASS&4\\
$  K  $      &  &$ 12.58\pm 0.03 $ &1998 Jan. 12& &$ 13.67\pm 0.04 $ & 1997 Dec. 18 && $ 12.90\pm 0.03 $ & 2000 Apr. 12& &2MASS&5\\
$  W1 $      &  &$ 11.07\pm 0.01 $ &2010 May~~01& &$ 11.34\pm 0.01 $ & 2010 Mar. 31 && $ 11.42\pm 0.02 $ & 2010 Jun~~07&  &WISE&5\\
$  W2 $      &  &$ 10.03\pm 0.01 $ &2010 May~~01& &$ ~9.86\pm 0.01 $ & 2010 Mar. 31 && $ 10.31\pm 0.02 $ & 2010 Jun~~07&  &WISE&5\\
$  W3 $      &  &$ ~7.33\pm 0.01 $ &2010 May~~01& &$ ~7.06\pm 0.01 $ & 2010 Mar. 31 && $ ~7.43\pm 0.02 $ & 2010 Jun~~07&  &WISE&5\\
$  W4 $      &  &$ ~4.80\pm 0.01 $ &2010 May~~01& &$ ~5.25\pm 0.03 $ & 2010 Mar. 31 && $ ~4.81\pm 0.02 $ & 2010 Jun~~07&  &WISE&5\\
$F475W$      &  &                  &			& &                  & 2005 May~~26 &&                   & 			   &  &HST/ACS&6\\
$F814W$      &  &                  &			& &                  & 2005 May~~26 &&                   & 			   &  &HST/ACS&6\\
%http://archive.stsci.edu/cgi-bin/mastpreview?mission=hst&dataid=J95W08010
%http://archive.stsci.edu/cgi-bin/mastpreview?mission=hst&dataid=J95W08030
%3 - 79 keV   &  &                  &            & &                  &              &&                   & 2014 Jul. 31&  &NuSTAR&6\\
%0.5 - 8 keV  &  &                  &2003 Apr. ~~& &                  &              &&                   & 2003 Apr. 30&  &$Chandra$&7\\
1100 - 6000 \AA&  &                 &2002 Apr. 27& &                  & 				&&                   & 			   &  &HST/STIS&7\\
%http://archive.stsci.edu/cgi-bin/mastpreview?mission=hst&dataid=O6G903010
3800 - 9200 \AA&  &                 &2005 Nov. 25& &                  & 2004 Mar. 14 &&                   & 2006 Mar. 24&  &SDSS&3\\
0.9 - 2.46 $\mu$m&&                 &2015 Mar. 11& &                  & 2014 Jan. 17 &&                   & 2014 Jan. 16&  &P200/TripleSpec& This work
%http://irsa.ipac.caltech.edu/images.html  MIR data
\enddata
\tablenotetext{References:}{(1) Morrissey et al. (2007); (2) York et al. (2000); (3) Abazajian et al. (2009); (4) Skrutskie et al. (2006); (5) Wright et al. (2010); (6) Urrutia et al. (2008); (7)Kuraszkiewicz et al.(2009)}
\tablenotetext{a:}{Flux in 7.5 arcsec diameter aperture, obtained from http://ned.ipac.caltech.edu}
\end{deluxetable}
%http://ned.ipac.caltech.edu/cgi-bin/datasearch?search_type=Photo_id&objid=102647&objname=2MASS%20J09184860%2B2117170&img_stamp=YES&hconst=73.0&omegam=0.27&omegav=0.73&corr_z=1&of=table

\begin{deluxetable}{c ccc c ccc c ccc}
\tabletypesize{\scriptsize}
\tablewidth{0pt}
\rotate
\tablenum{2}
\tablecaption{Broad Emission Lines Parameters and Balmer Decrement
\label{tab2} }
\tablehead{
\colhead{ }  &\multicolumn{3}{c}{SDSS J0918+2117} &\colhead{}& \multicolumn{3}{c}{SDSS J1113+1244} &\colhead{}& \multicolumn{3}{c}{SDSS J1227+3214}\\
\cline{2-4} \cline{6-8} \cline{10-12}\\
\hline
\colhead{Line} &\colhead{Centroid} & \colhead{FWHM} & \colhead{Flux} & \colhead{} &\colhead{Centroid} & \colhead{FWHM} & \colhead{Flux} &\colhead{} &\colhead{Centroid} & \colhead{FWHM} & \colhead{Flux}\\
\colhead{} &\colhead{(\AA)} & \colhead{(\kms)} & \colhead{($\rm 10^{-17}~erg~cm^{-2}$)} &\colhead{} &\colhead{((\AA))} & \colhead{(\kms)} & \colhead{($\rm 10^{-17}~erg~cm^{-2}$)} &\colhead{} &\colhead{\AA} & \colhead{(\kms)} & \colhead{($\rm 10^{-17}~erg~cm^{-2}$)} }
\startdata
H$\beta$  & 4862.62 & 1383 & $1763\pm 47$  & & 4862.60 & 1658 & $357\pm18$  & & 4862.74 & 1175 & $1069\pm 26$ \\
H$\alpha$ & 6564.02 & 1381 & $11652\pm 230$& & 6563.25 & 1832 & $3145\pm143$& & 6565.70 & 1244 & $10844\pm 294$\\
P$\alpha$ & 18759.37& 1123 & $2019\pm 307$ & &  -      &  -   &          -        & & 18758.55& 1094 & $3591\pm283$\\
\hline
Decrement &         & Value&   $E(B-V)$          & &         & Value&   $E(B-V)$         & &         & Value&   $E(B-V)$         \\
\hline
H$\alpha$/H$\beta$& &$6.61\pm0.22$&$0.84\pm0.04$ & &         &$8.81\pm0.59$&$1.15\pm0.07$&  &        &$10.14\pm0.38$&$1.30\pm0.04$ \\
P$\alpha$/H$\beta$& &$1.15\pm0.18$&$0.50\pm0.06$ & &         &  -   &      -             &  &        &$3.36\pm0.28$ &$0.94\pm0.03$
\enddata
\tablenotetext{Note:}{Adopting the multicomponent profile of the broad emission lines}
\end{deluxetable}

\clearpage

\figurenum{1}

\figurenum{2}

\figurenum{3}

\figurenum{4}

\figurenum{5}

\figurenum{6}

\figurenum{7}


\begin{thebibliography}{}
\bibitem[Abazajian et al.(2009)]{2009ApJS..182..543A} Abazajian, K.~N., Adelman-McCarthy, J.~K., Ag{\"u}eros, M.~A., et al.\ 2009, \apjs, 182, 543-558
\bibitem[Ai et al.(2013)]{2013AJ....145...90A} Ai, Y.~L., Yuan, W., Zhou, H., et al.\ 2013, \aj, 145, 90
\bibitem[Antonucci(1993)]{1993ARA&A..31..473A} Antonucci, R.\ 1993, \araa, 31, 473
\bibitem[Arav et al.(2001)]{2001ApJ...546..140A} Arav, N., Brotherton, M.~S., Becker, R.~H., et al.\ 2001, \apj, 546, 140
\bibitem[Bian \& Huang(2010)]{2010SCPMA..53S.256B} Bian, W., \& Huang, L.\ 2010, Science China Physics, Mechanics, and Astronomy, 53, 256
\bibitem[Boroson \& Green(1992)]{1992ApJS...80..109B} Boroson, T.~A., \& Green, R.~F.\ 1992, \apjs, 80, 109
\bibitem[Burtscher et al.(2013)]{2013A&A...558A.149B} Burtscher, L., Meisenheimer, K., Tristram, K.~R.~W., et al.\ 2013, \aap, 558, A149
\bibitem[Crenshaw et al.(2003)]{2003AJ....126.1690C} Crenshaw, D.~M., Kraemer, S.~B., \& Gabel, J.~R.\ 2003, \aj, 126, 1690
\bibitem[Cushing et al.(2004)]{2004PASP..116..362C} Cushing, M.~C., Vacca, W.~D., \& Rayner, J.~T.\ 2004, \pasp, 116, 362
\bibitem[Deo et al.(2006)]{2006AJ....132..321D} Deo, R.~P., Crenshaw, D.~M., \& Kraemer, S.~B.\ 2006, \aj, 132, 321
\bibitem[Dimitrijevi{\'c} et al.(2007)]{2007MNRAS.374.1181D} Dimitrijevi{\'c}, M.~S., Popovi{\'c}, L.~{\v C}., Kova{\v c}evi{\'c}, J., Da{\v c}i{\'c}, M., \& Ili{\'c}, D.\ 2007, \mnras, 374, 1181
\bibitem[Dong et al.(2005)]{2005ApJ...620..629D} Dong, X.-B., Zhou, H.-Y., Wang, T.-G., et al.\ 2005, \apj, 620, 629
\bibitem[Dong et al.(2008)]{2008MNRAS.383..581D} Dong, X., Wang, T., Wang, J., et al.\ 2008, \mnras, 383, 581
\bibitem[Dunn et al.(2010)]{2010ApJ...709..611D} Dunn, J.~P., Bautista, M., Arav, N., et al.\ 2010, \apj, 709, 611
\bibitem[Ferland \& Netzer(1983)]{1983ApJ...264..105F} Ferland, G.~J., \& Netzer, H.\ 1983, \apj, 264, 105
\bibitem[Ferland et al.(1998)]{1998PASP..110..761F} Ferland, G.~J., Korista, K.~T., Verner, D.~A., et al.\ 1998, \pasp, 110, 761
\bibitem[Fitzpatrick(1999)]{1999PASP..111....63} Fitzpatrick, E. L. 1999, PASP, 111, 63
\bibitem[Gaskell(1984)]{1984ApL....24...43G} Gaskell, C.~M.\ 1984, \aplett, 24, 43
\bibitem[Gaskell \& Ferland(1984)]{1984PASP...96..393G} Gaskell, C.~M., \& Ferland, G.~J.\ 1984, \pasp, 96, 393
\bibitem[Glikman et al.(2006)]{2006ApJ...640..579G} Glikman, E., Helfand, D.~J., \& White, R.~L.\ 2006, \apj, 640, 579
\bibitem[Glikman et al.(2007)]{2007ApJ...667..673G} Glikman, E., Helfand, D.~J., White, R.~L., et al.\ 2007, \apj, 667, 673
\bibitem[Glikman et al.(2012)]{2012ApJ...757...51G} Glikman, E., Urrutia, T., Lacy, M., et al.\ 2012, \apj, 757, 51
\bibitem[Goodrich(1989)]{1989ApJ...342..224G} Goodrich, R.~W.\ 1989, \apj, 342, 224
\bibitem[Grupe et al.(1999)]{1999A&A...350..805G} Grupe, D., Beuermann, K., Mannheim, K., \& Thomas, H.-C.\ 1999, \aap, 350, 805
\bibitem[Grupe \& Mathur(2004)]{2004ApJ...606L..41G} Grupe, D., \& Mathur, S.\ 2004, \apjl, 606, L41
\bibitem[Grupe et al.(2004)]{2004AJ....127..156G} Grupe, D., Wills, B.~J., Leighly, K.~M., \& Meusinger, H.\ 2004, \aj, 127, 156
\bibitem[Hewett \& Wild(2010)]{2010MNRAS.405.2302H} Hewett, P.~C., \& Wild, V.\ 2010, \mnras, 405, 2302
\bibitem[Ji et al.(2015)]{2015ApJ...800...56J} Ji, T., Zhou, H., Jiang, P., et al.\ 2015, \apj, 800, 56
\bibitem[Kishimoto et al.(2011)]{2011A&A...536A..78K} Kishimoto, M., H{\"o}nig, S.~F., Antonucci, R., et al.\ 2011, \aap, 536, A78
\bibitem[Komossa et al.(2008)]{2008ApJ...680..926K} Komossa, S., Xu, D., Zhou, H., Storchi-Bergmann, T., \& Binette, L.\ 2008, \apj, 680, 926-938
\bibitem[Krongold et al.(2001)]{2001AJ....121..702K} Krongold, Y., Dultzin-Hacyan, D., \& Marziani, P.\ 2001, \aj, 121, 702
\bibitem[Kuraszkiewicz et al.(2009)]{2009ApJ...692.1143K} Kuraszkiewicz, J., Wilkes, B.~J., Schmidt, G., et al.\ 2009, \apj, 692, 1143
\bibitem[LaMassa et al.(2016)]{2016ApJ...820...70L} LaMassa, S.~M., Ricarte, A., Glikman, E., et al.\ 2016, \apj, 820, 70
\bibitem[Leighly(1999)]{1999ApJS..125..317L} Leighly, K.~M.\ 1999, \apjs, 125, 317
\bibitem[Le{\'o}n Tavares et al.(2014)]{2014ApJ...795...58L} Le{\'o}n Tavares, J., Kotilainen, J., Chavushyan, V., et al.\ 2014, \apj, 795, 58
\bibitem[Leighly et al.(2011)]{2011ApJ...728...94L} Leighly, K.~M., Dietrich, M., \& Barber, S.\ 2011, \apj, 728, 94
\bibitem[Li et al.(2015)]{2015ApJ...812...99L} Li, Z., Zhou, H., Hao, L., et al.\ 2015, \apj, 812, 99
\bibitem[Liu et al.(2015)]{2015ApJS..217...11L} Liu, W.-J., Zhou, H., Ji, T., et al.\ 2015, \apjs, 217, 11
\bibitem[Liu et al.(2016)]{2016ApJ...822...64L} Liu, W.-J., Zhou, H.-Y., Jiang, N., et al.\ 2016, \apj, 822, 64
\bibitem[Markwardt(2009)]{2009ASPC..411..251M} Markwardt, C.~B.\ 2009, Astronomical Data Analysis Software and Systems XVIII, 411, 251
\bibitem[Marziani et al.(2001)]{2001ApJ...558..553M} Marziani, P., Sulentic, J.~W., Zwitter, T., Dultzin-Hacyan, D., \& Calvani, M.\ 2001, \apj, 558, 553
\bibitem[Mathur(2000)]{2000MNRAS.314L..17M} Mathur, S.\ 2000, \mnras, 314, L17
\bibitem[Monnier \& Millan-Gabet(2002)]{2002ApJ...579..694M} Monnier, J.~D., \& Millan-Gabet, R.\ 2002, \apj, 579, 694
\bibitem[Mor \& Trakhtenbrot(2011)]{2011ApJ...737L..36M} Mor, R., \& Trakhtenbrot, B.\ 2011, \apjl, 737, L36
\bibitem[Morrissey et al.(2007)]{2007ApJS..173..682M} Morrissey, P., Conrow, T., Barlow, T.~A., et al.\ 2007, \apjs, 173, 682
\bibitem[Netzer et al.(2007)]{2007ApJ...666..806N} Netzer, H., Lutz, D., Schweitzer, M., et al.\ 2007, \apj, 666, 806
\bibitem[Ohta et al.(2007)]{2007ApJS..169....1O} Ohta, K., Aoki, K., Kawaguchi, T., \& Kiuchi, G.\ 2007, \apjs, 169, 1
\bibitem[Olgu{\'{\i}}n-Iglesias et al.(2017)]{2017MNRAS.467.3712O} Olgu{\'{\i}}n-Iglesias, A., Kotilainen, J.~K., Le{\'o}n Tavares, J., Chavushyan, V., \& A{\~n}orve, C.\ 2017, \mnras, 467, 3712
\bibitem[Ohta et al.(2007)]{2007ApJS..169....1O} Ohta, K., Aoki, K., Kawaguchi, T., \& Kiuchi, G.\ 2007, \apjs, 169, 1
\bibitem[Olgu{\'{\i}}n-Iglesias et al.(2017)]{2017MNRAS.467.3712O} Olgu{\'{\i}}n-Iglesias, A., Kotilainen, J.~K., Le{\'o}n Tavares, J., Chavushyan, V., \& A{\~n}orve, C.\ 2017, \mnras, 467, 3712
\bibitem[Osterbrock \& Pogge(1985)]{1985ApJ...297..166O} Osterbrock, D.~E., \& Pogge, R.~W.\ 1985, \apj, 297, 166
\bibitem[Osterbrock \& Pogge(1987)]{1987ApJ...323..108O} Osterbrock, D.~E., \& Pogge, R.~W.\ 1987, \apj, 323, 108
\bibitem[Pan et al.(2017)]{2017ApJ...835..218P} Pan, X., Zhou, H., Ge, J., et al.\ 2017, \apj, 835, 218
\bibitem[Peng et al.(2010)]{2010AJ....139.2097P} Peng, C.~Y., Ho, L.~C., Impey, C.~D., \& Rix, H.-W.\ 2010, \aj, 139, 2097
\bibitem[Peterson(1997)]{1997iagn.book.....P} Peterson, B.~M.\ 1997, An introduction to active galactic nuclei, Publisher: Cambridge, New York Cambridge University Press, 1997 Physical description xvi, 238 p.~ISBN 0521473489,
\bibitem[Polletta et al.(2007)]{2007ApJ...663...81P} Polletta, M., Tajer, M., Maraschi, L., et al.\ 2007, \apj, 663, 81
\bibitem[Pounds \& Wilkes(2007)]{2007MNRAS.380.1341P} Pounds, K.~A., \& Wilkes, B.~J.\ 2007, \mnras, 380, 1341
\bibitem[Puchnarewicz et al.(1992)]{1992MNRAS.256..589P} Puchnarewicz, E.~M., Mason, K.~O., Cordova, F.~A., et al.\ 1992, \mnras, 256, 589
%\bibitem[Rayner et al.(2009)]{2009ApJS..185..289R} Rayner, J.~T., Cushing, M.~C., \& Vacca, W.~D.\ 2009, \apjs, 185, 289
\bibitem[Rovilos et al.(2009)]{2009A&A...502...85R} Rovilos, E., Georgantopoulos, I., Tzanavaris, P., et al.\ 2009, \aap, 502, 85
\bibitem[Schlegel et al.(1998)]{1998ApJ...500..525S} Schlegel, D.~J., Finkbeiner, D.~P., \& Davis, M.\ 1998, \apj, 500, 525
\bibitem[Skrutskie et al.(2006)]{2006AJ....131.1163S} Skrutskie, M.~F., Cutri, R.~M., Stiening, R., et al.\ 2006, \aj, 131, 1163
\bibitem[Storey \& Zeippen(2000)]{2000MNRAS.312..813S} Storey, P.~J., \& Zeippen, C.~J.\ 2000, \mnras, 312, 813
\bibitem[Stoughton et al.(2002)]{2002AJ....123..485S} Stoughton, C., Lupton, R.~H., Bernardi, M., et al.\ 2002, \aj, 123, 485
\bibitem[Sun et al.(2017)]{2017ApJ...838...88S} Sun, L., Zhou, H., Ji, T., et al.\ 2017, \apj, 838, 88
%\bibitem[Sun et al.(2017)]{2017arXiv170302686S} Sun, L., Zhou, H., Ji, T., et al.\ 2017, arXiv:1703.02686
\bibitem[Taniguchi et al.(1999)]{1999astro.ph.10036T} Taniguchi, Y., Murayama, T., \& Nagao, T.\ 1999, arXiv:astro-ph/9910036
\bibitem[Tuthill et al.(2001)]{2001Natur.409.1012T} Tuthill, P.~G., Monnier, J.~D., \& Danchi, W.~C.\ 2001, \nat, 409, 1012
\bibitem[Veilleux \& Osterbrock(1987)]{1987ApJS...63..295V} Veilleux, S., \& Osterbrock, D.~E.\ 1987, \apjs, 63, 295
\bibitem[Urrutia et al.(2008)]{2008ApJ...674...80U} Urrutia, T., Lacy, M., \& Becker, R.~H.\ 2008, \apj, 674, 80-96
\bibitem[Urrutia et al.(2012)]{2012ApJ...757..125U} Urrutia, T., Lacy, M., Spoon, H., et al.\ 2012, \apj, 757, 125
\bibitem[Vacca et al.(2003)]{2003PASP..115..389V} Vacca, W.~D., Cushing, M.~C., \& Rayner, J.~T.\ 2003, \pasp, 115, 389
\bibitem[Vanden Berk et al.(2001)]{2001AJ....122..549V} Vanden Berk, D.~E., Richards, G.~T., Bauer, A., et al.\ 2001, \aj, 122, 549
\bibitem[V{\'e}ron-Cetty et al.(2004)]{2004A&A...417..515V} V{\'e}ron-Cetty, M.-P., Joly, M., \& V{\'e}ron, P.\ 2004, \aap, 417, 515
\bibitem[Wang et al.(1996)]{1996A&A...309...81W} Wang, T., Brinkmann, W., \& Bergeron, J.\ 1996, \aap, 309, 81
\bibitem[Weingartner \& Draine(2001)]{2001ApJ...548..296W} Weingartner, J.~C., \& Draine, B.~T.\ 2001, \apj, 548, 296
\bibitem[Wilson et al.(2004)]{2004SPIE.5492.1295W} Wilson, J.~C., Henderson, C.~P., Herter, T.~L., et al.\ 2004, \procspie, 5492, 1295
\bibitem[Wright et al.(2010)]{2010AJ....140.1868W} Wright, E.~L., Eisenhardt, P.~R.~M., Mainzer, A.~K., et al.\ 2010, \aj, 140, 1868
\bibitem[Xu et al.(2012)]{2012AJ....143...83X} Xu, D., Komossa, S., Zhou, H., et al.\ 2012, \aj, 143, 83
\bibitem[York et al.(2000)]{2000AJ....120.1579Y} York, D.~G., Adelman, J., Anderson, J.~E., Jr., et al.\ 2000, \aj, 120, 1579
\bibitem[Zhang et al.(2011)]{2011ApJ...737...71Z} Zhang, K., Dong, X.-B., Wang, T.-G., \& Gaskell, C.~M.\ 2011, \apj, 737, 71
\bibitem[Zhang et al.(2017)]{2017ApJ...836...86Z} Zhang, S., Zhou, H., Shi, X., et al.\ 2017, \apj, 836, 86
\bibitem[Zhang et al.(2009)]{2009ApJ...699..281Z} Zhang, W.~M., Soria, R., Zhang, S.~N., Swartz, D.~A., \& Liu, J.~F.\ 2009, \apj, 699, 281
\bibitem[Zhou et al.(2006)]{2006ApJS..166..128Z} Zhou, H., Wang, T., Yuan, W., et al.\ 2006, \apjs, 166, 128
\end{thebibliography}
\end{document}